# JGR Atmospheres

AGU100 ADVANCING EARTH AND SPACE SCIENCE

**Key Points:**
- TGFs detected by AGILE with counts energy larger than 40 MeV are compatible with RREA
- Data show no evidence of a fine time structure of TGFs on microsecond time scale.
- TGFs simultaneous to lightning suggest that the AGILE TGF sample can be significantly increased

# On the High-Energy Spectral Component and Fine Time Structure of Terrestrial Gamma Ray Flashes

M. Marisaldi[1,2] , M. Galli[3], C. Labanti[2], N. Østgaard[1] , D. Sarria[1], S. A. Cummer[4] , F. Lyu[4] , A. Lindanger[1] , R. Campana[2], A. Ursi[5], M. Tavani[5], F. Fuschino[2], A. Argan[6], A. Trois[7], C. Pittori[8] , and F. Verrecchia[8]

[1]Birkeland Centre for Space Science, Department of Physics and Technology, University of Bergen, Bergen, Norway, [2]INAF-OAS Bologna, Bologna, Italy, [3]ENEA, Bologna, Italy, [4]Electrical and Computer Engineering Department, Duke University, Durham, NC, USA, [5]INAF-IAPS Roma, Rome, Italy, [6]INAF, Rome, Italy, [7]INAF-Osservatorio Astronomico di Cagliari, Selargius (CA), Italy, [8]Space Science Data Center-Agenzia Spaziale Italiana, Rome, Italy

**Abstract** Terrestrial gamma ray flashes (TGFs) are very short bursts of gamma radiation associated to thunderstorm activity and are the manifestation of the highest-energy natural particle acceleration phenomena occurring on Earth. Photon energies up to several tens of megaelectronvolts are expected, but the actual upper limit and high-energy spectral shape are still open questions. Results published in 2011 by the AGILE team proposed a high-energy component in TGF spectra extended up to ≈100 MeV, which is difficult to reconcile with the predictions from the Relativistic Runaway Electron Avalanche (RREA) mechanism at the basis of many TGF production models. Here we present a new set of TGFs detected by the AGILE satellite and associated to lightning measurements capable to solve this controversy. Detailed end-to-end Monte Carlo simulations and an improved understanding of the instrument performance under high-flux conditions show that it is possible to explain the observed high-energy counts by a standard RREA spectrum at the source, provided that the TGF is sufficiently bright and short. We investigate the possibility that single high-energy counts may be the signature of a fine-pulsed time structure of TGFs on time scales ≈4 μs, but we find no clear evidence for this. The presented data set and modeling results allow also for explaining the observed TGF distribution in the (Fluence × duration) parameter space and suggest that the AGILE TGF detection rate can almost be doubled.

## 1. Introduction

Terrestrial gamma ray flashes (TGFs) are very short (typically submillisecond in duration) bursts of gamma radiation associated to thunderstorms and lightning activity (Briggs et al., 2010; Dwyer et al., 2012; Fishman, 1994; Marisaldi et al., 2010; Smith, 2005). TGFs represent the manifestation of the most energetic natural particle acceleration processes occurring on Earth within thundercloud electric fields. Photon energies up to several tens of megaelectronvolts have been reported, but the exact maximum energy that TGF photons can reach has not been clearly assessed yet. This is mostly due to the energy range of current TGF detectors which is limited to few tens of megaelectronvolts and therefore makes all counts with higher energy to be registered in the overload channel without accurate energy information. Moreover, an accurate measurement of photon energy in this range is difficult, since photon interaction cross section in typical detector materials is dominated by electron-positron pair production and a large amount of detecting material is required for a full energy measurement.

One of the most acknowledged physical processes thought to be at the basis of TGF production is the Relativistic Runaway Electron Avalanche (RREA) process (Gurevich et al., 1992), possibly enhanced by the Relativistic Feedback mechanism (Dwyer, 2003, 2012). In this scenario, the resulting TGF source photon spectrum is basically a power law with exponential cutoff with *e*-folding energy of ≈7.3 MeV; therefore, it is difficult to account for photon energies larger than 30–40 MeV. Cumulative spectra of TGFs detected by RHESSI and AGILE (Dwyer & Smith, 2005; Marisaldi et al., 2014) proved to be compliant with this expectation. However, the use of cumulative spectra itself is questionable because all effects due to atmospheric absorption from different source regions and direction-dependent detector response are smeared out and mixed together. Single photon maximum energy was reported by RHESSI, AGILE, and Fermi teams as larger than 20, 43, and 38 MeV, respectively (Briggs et al., 2010; Marisaldi et al., 2010; Smith, 2005). TGF detection by the AGILE Gamma Ray Imaging Detector (Marisaldi et al., 2010), sensitive above 20 MeV, indicates that the TGF spectral component in the tens of megaelectronvolts range is significant, although the energy resolution of the instrument close to the detection threshold is not sufficient to clearly assess the maximum photon energy. A systematic attempt at spectral fitting of individual TGFs detected by Fermi is reported in Mailyan et al. (2016). Out of the 46 TGFs analyzed, 5 of them show a poor fit because of excess counts at high energy. The authors state that it is not clear whether this is due to deviations of the source spectra from RREA predictions or underestimation of instrumental effects such as pulse pileup. In fact, due to the very high TGF photon flux, instrumental effects significantly affect the measurements from all TGF-observing instruments (Briggs et al., 2010; Gjesteland et al., 2010; Grefenstette et al., 2009; Marisaldi et al., 2014) and must be carefully accounted for when dealing with TGF intensity and energy spectrum. In 2010 the AGILE team, including some of the authors of this paper, reported the detection of TGFs by the minicalorimeter (MCAL) instrument onboard AGILE with photon energies up to 100 MeV (Tavani et al., 2011), with a significant deviation from predictions by the RREA model. These results triggered significant theoretical efforts for their interpretation, (e.g., Celestin et al., 2012, 2015; Luque, 2014). We also note that gamma ray differential energy spectra extended up to 100 MeV have been reported for long-lasting Thunderstorm Ground Enhancements (TGEs) observed on ground (Chilingarian et al., 2013). However, independent confirmation of these findings were never obtained, basically because of the energy range of other space-based TGF observing instruments being limited to 40 MeV. A thorough understanding of the TGF emission spectrum in the tens of megaelectronvolt range is particularly relevant also for the quantitative assessment of neutrons and radioactive isotopes production by photonuclear reactions (Babich & Roussel-Dupre, 2007; Babich et al., 2014; Bowers et al., 2017; Carlson et al., 2010; Enoto et al., 2017; Tavani et al., 2013).

The discovery of TGFs simultaneous (within few hundreds of microseconds) to lightning sferics detected by ground-based lightning detection networks (Connaughton et al., 2010, 2013) allowed to use only the association to lightning itself for TGF identification, provided a minimum number of counts are present, without the need for additional selection criteria. In other words, if any cluster of counts is observed in close time association to a lightning, the probability of chance association is remote and we can be reasonably sure it is a TGF, regardless of all its other properties (Albrechtsen et al., 2019; Østgaard et al., 2015). A set of events firmly associated to lightning sferics would provide a reliable sample of TGFs unbiased by selection criteria based on gamma ray data only, and would provide a test bench to confirm or disprove the existence of photons with energy higher than 40 MeV in TGF spectra. With this motivation, we searched for clusters of counts associated with lightning, without introducing any additional selection criteria. However, no simultaneous association of AGILE MCAL events to lightning sferics were found before 23 March 2015. This was due to the suppression of the detection of short TGFs due to the dead time induced by the anticoincidence (AC) shield surrounding the MCAL instrument (Marisaldi et al., 2014). In turn, the chance of association to lightning sferics detected by ground-based lightning detection networks strongly decrease with increasing TGF duration (Connaughton et al., 2013). Starting from 23 March 2015, the AC veto signal was inhibited for MCAL, resulting in a tenfold increase in TGF detection rate. Between 23 March and 24 June 2015, a total of 279 TGFs have been recorded by AGILE in this enhanced configuration (Marisaldi et al., 2015) using the standard selection criteria described in Marisaldi et al. (2014). Among them, 39 events are associated with a lightning sferic detected by the World-Wide Lightning Location Network (WWLLN) within 200 μs, when the propagation time from source to satellite is accounted for. For the same reference period, a TGF search based on simultaneous association to WWLLN sferics only resulted in the identification of 84 events, 28 of which exhibiting maximal count energy above 30 MeV. These events were rejected by the previously applied selection criteria. This data set provides the test bench we need to investigate the TGF maximal photon energy. We stress the fact that, although the claim for a population of high-energy TGFs dates back to 2011 (Tavani et al., 2011), we were not able to pursue this analysis until a data set with simultaneous lightning association was available, that is, after the major configuration change of 23 March 2015.

In this work we study the properties of this WWLLN-identified sample with respect to previous observations. Then we describe an end-to-end simulation frame used to interpret the observations by taking correctly into account the TGF source spectrum, photon propagation to satellite altitude, the detector energy response, and the behavior of the front-end electronics. Finally, we discuss the results regarding implications on AGILE

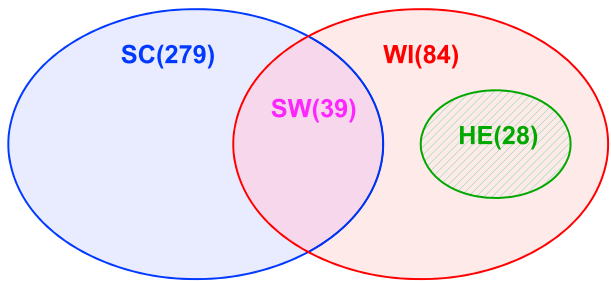

**Figure 1.** Schematic representation of the data sets used in this paper. SC: selection criteria data set. WI: WWLLN-identified data set. SW: SC subset with WWLLN identification. HE: WWLLN-identified events with high-energy (>30 MeV) counts. Numbers in parenthesis indicate the number of events in each data set. WWLLN = World-Wide Lightning Location Network.

TGF detection rate, fluence and duration distribution, the TGF high-energy spectral component, and fine time structure at the source.

## 2. High-Energy Data Set Characteristics

We scanned the AGILE MCAL data for the period 23 March to 24 June 2015 in search for associations between count clusters and lightning sferics detected by the WWLLN network. Data for the period after 1 July 2015 is affected by a degradation of the AGILE time performance due to a failure of the onboard GPS. We define a cluster as a minimum set of six counts detected in a 300-μs time window, which is the minimum requirement needed to trigger the MCAL onboard logic and enable data acquisition. We require a maximum time separation of ±500 μs between the sferic and the cluster, after correcting for light travel time from lightning location to the satellite. No additional selection criteria are introduced. The search resulted in 84 events associated to WWLLN sferics, hereafter the *WWLLN-identified* (WI) data set. The TGF data set including 279 events obtained by means of selection criteria and described in Marisaldi et al. (2015) is hereafter referred to as the *selection criteria* (SC) data set. All the 39 events in the SC data set associated to WWLLN lightning are identified also by the current analysis and included in the WI data set: these will be referred to as the *selection criteria and WWLLN* (SW) data set. Twenty-eight events of the WI data set exhibit at least one count with reconstructed energy larger than 30 MeV: these events, hereafter the *high-energy* (HE) data set, are the core target of this paper. Hereafter we will also indicate with the term *high-energy count* a count with measured energy larger than 30 MeV. Figure 1 shows a schematic representation of the data sets used throughout this paper, and colors on plots will also correspond to the same data set, accordingly.

AGILE has no onboard sources for energy calibration. Moreover, calibration in the tens of megaelectronvolts range is a nontrivial process. This is achieved for MCAL using galactic cosmic rays as calibration sources. The spatial segmentation of the MCAL detector allows the topological identification of cosmic ray tracks that ultimately allow the measurement of the specific energy loss for hydrogen and helium nuclei and compare them

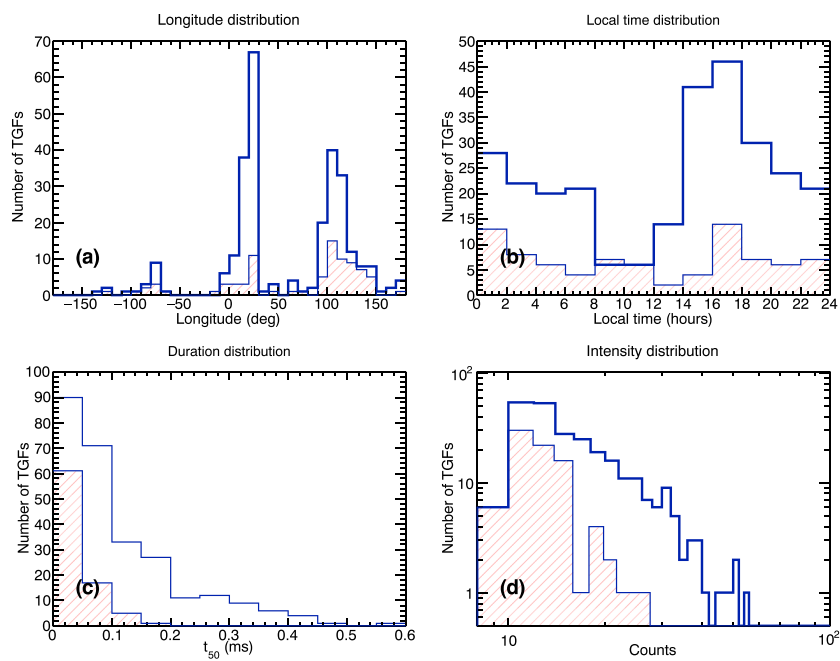

**Figure 2.** (a) Longitude, (b) local time, (c) duration, and (d) number of counts distributions for the selection criteria (blue) and WWLLN-identified (red hatches) data sets. WWLLN = World-Wide Lightning Location Network; TGFs = terrestrial gamma ray flashes.

**Table 1**
*TGF HE Sample Main Characteristics*

| Id | Date[a] (UT) | Lon$_A$[b] (°) | Lat$_A$[c] (°) | I[d] | $t_{50}$[e] (µs) | $E_{MAX}$[f] (MeV) | Lon$_W$[g] (°) | Lat$_W$[h] (°) | $\Delta t_W$[i] (µs) | D[j] (km) | $\theta$[k] (°) | $\phi$[l] (°) |
|---|---|---|---|---|---|---|---|---|---|---|---|---|
| 1 | 2015-6-11T16:18:16.985339 | 172.14 | −2.30 | 6.8 | 108 | 1879 | 167.37 | −3.39 | 164 | 545 | 121 | 338 |
| 2 | 2015-4-26T10:54:24.413659 | −7.71 | −2.01 | 8.7 | 84 | 1397 | −2.54 | −1.64 | 291 | 577 | 133 | 202 |
| 3 | 2015-4-20T14:12:27.607643 | 129.30 | 2.43 | 8.7 | 108 | 697 | 135.70 | −4.89 | −265 | 1082 | 85 | 3 |
| 4 | 2015-6-1T02:00:14.338701 | 102.33 | −1.95 | 9.7 | 84 | 607 | 98.89 | −2.20 | 120 | 384 | 99 | 244 |
| 5 | 2015-5-8T15:47:59.942848 | 23.86 | 0.99 | 12.9 | 62 | 528 | 24.25 | −1.76 | 48 | 310 | 130 | 197 |
| 6 | 2015-4-12T20:50:36.535486 | 73.73 | 2.08 | 8.7 | 151 | 323 | 88.58 | 9.95 | 201 | 1857 | 90 | 123 |
| 7 | 2015-5-26T08:15:27.310010 | 141.29 | −1.75 | 9.8 | 57 | 172 | 139.29 | −2.95 | 64 | 259 | 22 | 130 |
| 8 | 2015-4-25T10:10:42.663278 | 101.29 | 1.39 | 9.0 | 18 | 159 | 98.82 | 1.78 | 118 | 278 | 67 | 22 |
| 9 | 2015-6-11T16:04:14.697162 | 122.16 | −0.65 | 13.0 | 26 | 158 | 119.38 | 0.35 | 35 | 329 | 80 | 46 |
| 10 | 2015-4-21T05:21:58.090701 | 132.11 | −1.70 | 11.0 | 22 | 113 | 131.57 | −1.74 | 107 | 61 | 47 | 290 |
| 11 | 2015-4-11T10:52:19.591390 | 6.47 | −0.34 | 13.7 | 85 | 90 | 8.81 | 1.17 | 225 | 310 | 135 | 248 |
| 12 | 2015-4-4T11:12:30.780687 | 141.90 | −2.46 | 11.8 | 51 | 79 | 140.43 | −3.97 | 94 | 235 | 114 | 66 |
| 13 | 2015-3-31T14:35:18.935832 | 130.78 | −2.46 | 11.7 | 70 | 70 | 133.16 | −2.54 | 102 | 265 | 102 | 56 |
| 14 | 2015-5-19T09:34:02.818700 | 106.24 | −2.11 | 10.0 | 26 | 67 | 104.49 | −2.08 | 61 | 195 | 55 | 10 |
| 15 | 2015-5-1T02:29:30.522607 | 108.00 | −1.03 | 9.0 | 16 | 62 | 105.64 | −1.50 | −10 | 268 | 78 | 258 |
| 16 | 2015-5-10T11:26:51.004225 | 4.11 | 2.47 | 12.8 | 57 | 58 | 3.98 | 4.42 | 4 | 218 | 79 | 229 |
| 17 | 2015-5-20T14:57:07.442492 | −84.94 | 1.61 | 8.0 | 18 | 56 | -86.88 | 2.12 | 95 | 223 | 97 | 244 |
| 18 | 2015-5-21T17:06:30.955043 | 112.71 | 0.79 | 8.9 | 31 | 54 | 111.07 | 0.74 | 31 | 183 | 96 | 44 |
| 19 | 2015-5-21T09:28:13.853865 | −81.94 | 2.26 | 11.5 | 112 | 45 | −77.37 | 4.69 | 55 | 575 | 99 | 75 |
| 20 | 2015-5-23T01:02:07.902862 | 105.87 | 1.51 | 11.9 | 26 | 45 | 104.61 | 1.17 | 18 | 146 | 129 | 297 |
| 21 | 2015-4-9T05:07:00.052780 | 27.35 | 2.40 | 10.0 | 16 | 43 | 27.41 | 0.99 | 4 | 157 | 139 | 203 |
| 22 | 2015-6-5T05:12:20.006374 | 94.69 | −1.93 | 9.0 | 18 | 41 | 94.44 | −2.22 | 224 | 43 | 75 | 250 |
| 23 | 2015-6-5T18:43:24.888233 | 108.13 | 0.44 | 9.0 | 16 | 40 | 108.45 | −1.52 | −14 | 222 | 32 | 69 |
| 24 | 2015-5-5T09:12:52.047623 | 127.93 | 1.80 | 15.8 | 61 | 39 | 129.49 | 2.08 | 170 | 177 | 27 | 217 |
| 25 | 2015-4-12T20:56:56.613399 | 96.26 | 2.44 | 8.0 | 18 | 38 | 94.41 | 2.72 | −18 | 209 | 167 | 120 |
| 26 | 2015-4-8T07:12:19.336028 | 20.33 | 2.14 | 9.0 | 16 | 36 | 20.87 | −0.41 | −7 | 290 | 51 | 222 |
| 27 | 2015-5-26T19:52:37.161157 | 107.78 | 2.43 | 12.0 | 26 | 31 | 105.73 | 0.82 | 41 | 290 | 55 | 183 |
| 28 | 2015-6-1T12:09:01.001964 | 114.24 | 1.58 | 8.9 | 36 | 30 | 113.70 | 4.11 | −23 | 288 | 129 | 116 |

*Note.* TGF = terrestrial gamma ray flash; HE = high energy; WWLLN = World-Wide Lightning Location Network. Dates are formatted as year-month-day. [a]Event identifier. [b]Start time of the TGF (UTC), defined as the time of arrival of the first count. [c]AGILE subsatellite point longitude. [d]AGILE subsatellite point latitude. [e]Number of counts, after maximum likelihood fit with a Gaussian time profile. [f] The $t_{50}$ after maximum likelihood fit with a Gaussian time profile. [g]Maximum count energy. [h]Associated WWLLN event longitude. [i]Associated WWLLN event latitude. [j]Associated WWLLN event time delay with respect to TGF start time after correction for propagation time. [k]Associated WWLLN event distance to subsatellite point. [l]Incidence angle (zenith) of the incoming photons with respect to the satellite pointing direction. [m]Azimuth angle of the incoming photons with respect to the satellite reference frame.

to Monte Carlo simulations. Following this approach we can state that the MCAL energy reconstruction in the tens of megaelectronvolts range is accurate within 20% (1$\sigma$), including systematic errors.

Figure 2 shows the main parameter (longitude, local time, duration, and intensity) distributions for the SC and the WI sample. TGF duration throughout this paper is described by the $t_{50}$ parameter, defined as the time interval that includes the central 50% of the counts in a transient event (Koshut et al., 1996). Given the low number of counts in a transient, duration and intensity are derived by an unbinned maximum likelihood fitting procedure assuming a Gaussian time profile and a constant background, as described in detail in Marisaldi et al. (2014). The TGF duration is then calculated as $t_{50} = 1.349\sigma$, where $\sigma$ is the standard deviation of the Gaussian model. Differences in longitude and local time distributions are affected by the differences in WWLLN detection efficiency with respect to geographical region (driven by the geographic distribution of the receiving stations) and local time (driven by the day/night ionospheric asymmetry affecting the radio waves propagation properties). Differences in the duration distributions are due to the bias toward short durations for TGFs associated to sferics explained in details in Connaughton et al. (2013) and Dwyer and

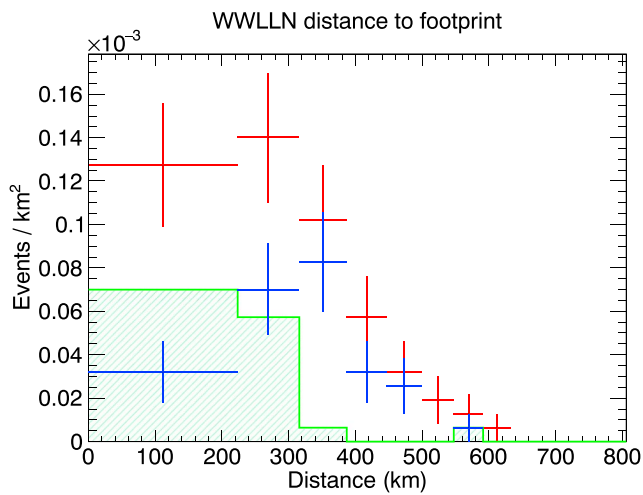

**Figure 3.** Distribution of the distance between the WWLLN location and the satellite footprint for the WI sample (red data points), the SW sample (blue data points), and the HE sample (green hatched histogram). Error bars correspond to the square root of the bin content. WWLLN = World-Wide Lightning Location Network; WI = WWLLN-identified; SW = SC subset with WWLLN identification; SC = selection criteria; HE = WWLLN-identified events with high energy.

Cummer (2013). The difference in the intensity, that is, the number of counts, distributions can also be ascribed to the difference in duration, since shorter AGILE events typically exhibit lower intensity (Marisaldi et al., 2015), as it will be discussed in section 4.2.

Table 1 shows the main characteristics of the 28 events belonging to the HE data set. The events are identified by a numerical Id which will be used throughout this work and are ordered according to decreasing maximum count energy ($E_{MAX}$). An extended version of the table, including links to all light curves and energy versus time scatter plots, for the WI data set can be accessed at this URL (http://www.ssdc.asi.it/mcalwtgfcat/). The events with Id 1–6 present extremely high count energy, ranging from 1.9 GeV down to 323 MeV. After a close look at the light curves and counts topology, we regard these as spurious events and we disregard them from subsequent analysis. High-energy background counts are due to cosmic ray particles, and their rate in MCAL depends on magnetic latitude. Based on averaged background observations, a good upper estimate is about 180 counts/s with reconstructed energy above 30 MeV. The probability of having a background count above 30 MeV in a 0.5-ms time window centered around the TGF time is therefore 0.09. If we consider this as a binomial process, the expected average number of positive results out of 28 trials (the total number of events in the high-energy sample) is 2.5, while we have at least six. This may be due to the fact that very high energy deposits in MCAL can be associated to instrumental effects resulting in the collection of two or more counts closely separated in time. This would in turn bias our requirement of minimum six counts per cluster, making this condition satisfied even for a smaller number of independent counts. Events #4 and #5 present the high-energy count well separated in time from the main TGF. These are presumably regular TGFs with maximum energy lower than 30 MeV contaminated by a background cosmic ray. These are also the only two of these spurious events with distance to the satellite footprint lower than 400 km. In particular, events #3 and #6 present distance to the satellite footprint larger than 1,000 km. These are most likely chance WWLLN associations, as real TGF photons at these distances would be almost completely absorbed in the atmosphere or Comptonized to energies lower than 500 keV (Hazelton et al., 2009; Østgaard et al., 2008; Smith et al., 2016).

Figure 3 shows the distribution of the distance between the WWLLN location and the satellite footprint for the WI, SW, and HE data sets, excluding the spurious events with Id 1–6 described above. Distance bin size has been chosen so that the surface area corresponding to each distance bin is constant and equal to $1.57 \cdot 10^5$ km². Therefore, the distributions represent the TGF surface density, apart from a normalization factor due to satellite exposure time. Implications for the AGILE TGF detection rate are discussed in section 4.1.

In the following we will focus on three case studies: events #10, #14, and #24 from Table 1. This choice has been made because their maximum count energy spans the range between 39 MeV (event #24, close to the plausible maximum energy expected from a RREA process) to 113 MeV (event #10, not compatible with RREA). Moreover the incidence zenith angle $\theta$ for these events is smaller than 55° so that the photons reached the detector without scattering in the spacecraft, making the detector response matrix more reliable. We note that many other events in Table 1 satisfy this last condition. Concerning the lower maximum energy events, we regard events #23, #24, #26, and #27 as equivalent, so our choice for #24 was arbitrary. For the intermediate energy range, only event #14 was a viable choice. Concerning the highest maximum energy, we regard event #10 as representative of the class of events that dominate the high-energy power law component of the cumulative spectrum presented in Tavani et al. (2011) and we decided not to choose as case study the extreme case represented by #7. Figure 4 shows the light curves and count energy versus time plots for these events. In the top plots, the cyan curve is the maximum likelihood fit to the data assuming a Gaussian time profile, the magenta vertical line indicates the occurrence time of the WWLLN match corrected for light propagation time to the spacecraft, assuming a 15-km source altitude (Dwyer & Smith, 2005; Dwyer et al., 2012).

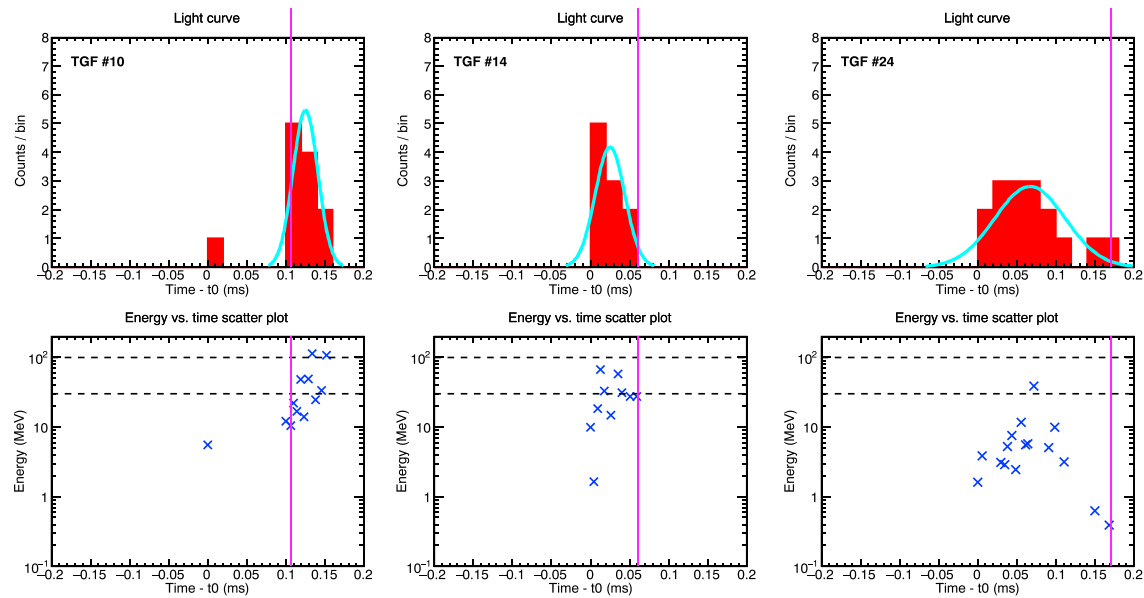

**Figure 4.** Light curve (top row) and count energy versus time (bottom row) for events #10, #14, and #24. In the top plots, the cyan curve is the maximum likelihood fit to the data assuming a Gaussian model. In the bottom plots the horizontal dashed lines mark 30- and 100-MeV energy levels. The magenta vertical lines indicate the occurrence time of the WWLLN match. The parameter $t_0$ is the start time of the TGF reported in Table 1.

Event #10 consists of 11 counts recorded in about 50 µs, two counts have energy above 100 MeV, and all counts but the first one have energy above 10 MeV. Such a very energetic spectrum is difficult to reconcile with expected production models. The time separation between consecutive counts is always close to 4 µs, which is a lower limit set by the front-end electronics (FEE) design, see section 3.3. This means that, however high the incoming photon flux can be, no counts closer in time than 4 µs can be recorded, resulting in a maximum detectable flux of about 250 kHz. Photon signals in the detectors are then combined by the FEE in a nontrivial way to produce the measured counts, as described in section 3.3. We regard it as extremely unlikely that the incoming flux for this event, as well as for most of the others included in the sample, is so finely tuned to the maximum detectable rate of the instrument. Therefore, we assume that the true flux is higher than this maximum detectable rate. This gives a strong indication that the observations are significantly affected by instrumental effects.

## 3. Monte Carlo Modeling

### 3.1. Method

We soon realized that these events force the instrument to work in conditions of extremely high count rate, well above the design specifications. Therefore, a detailed understanding of the instrument analog and digital FEE is mandatory to properly interpret the measurements. Given the complexity of the FEE design, there is no simple reliable algorithm to map an observed pattern back to its parent physical photon pattern; therefore, we must use numerical simulations of the FEE with a forward folding approach, as described in this section. The instrument FEE model is described in section 3.3. We use this model to explore the TGF fluence and time profile parameter space in order to identify plausible scenarios compatible with the observations. For these simulations we follow a more detailed approach with respect to that already exploited in Marisaldi et al. (2014, 2015). The current approach is outlined in the following:

1. We model a RREA spectrum produced at 15-km altitude and use the GEANT4 toolkit (Agostinelli et al., 2003; Allison et al., 2006) to propagate the photons at satellite altitude for the specific viewing geometries corresponding to the case studies introduced above, see section 3.2.

2. We use the simulated spectra at satellite altitude obtained in step 1 as input to the Monte Carlo simulation code that includes the full mass model of the satellite. The appropriate incoming direction for the case studies is taken into account. The output of this step provides the actual hits in the detector (number of hit detector units, position, and energy deposition in each detector unit).

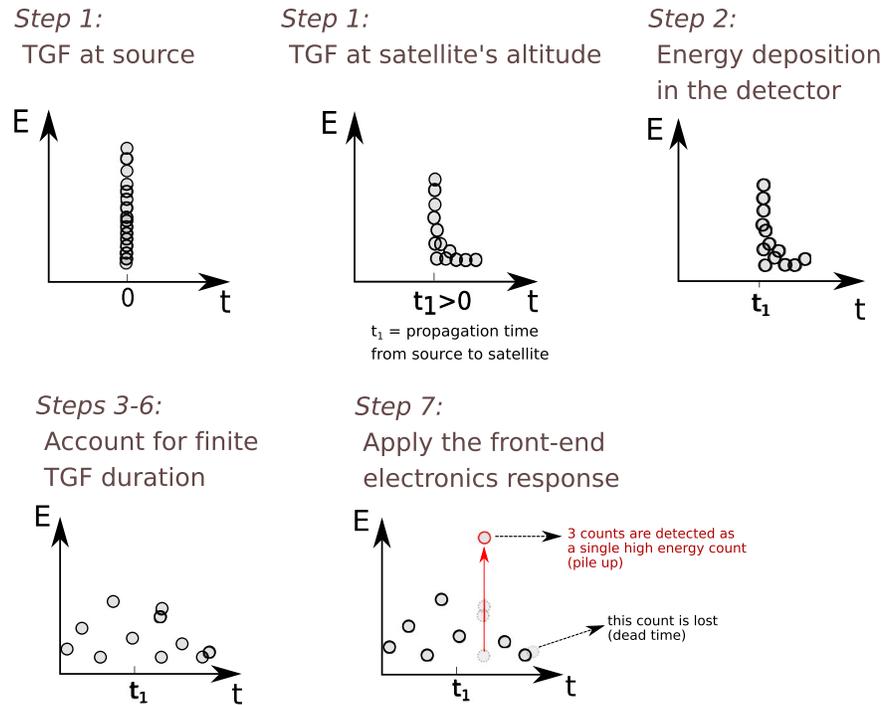

**Figure 5.** Schematic representation of the simulation flow. Step numbers refer to the numbered items in section 3.

3. We consider a TGF as parameterized by two main observables: its fluence at satellite altitude $F$ and its intrinsic duration expressed in terms of $t_{50}$. We assume here a Gaussian time profile for the intrinsic duration. Note that this duration is combined with the energy-dependent time spread dependent on transport through the atmosphere and observation geometry, obtained as output of step 1 and discussed in section 3.2.

4. We generate $10^6$ simulated TGFs uniformly distributed in the parameter space given by (0.05 cm$^{-2}$ < $F$ < 1.95 cm$^{-2}$) × (0.001 ms < $t_{50}$ < 0.5 ms).

5. For each simulated TGF, the expected number of counts in MCAL is defined according to $F$ and the average effective area given by simulation results; the time series of the counts were randomly extracted according to a Gaussian time profile with $\sigma = 0.74 \, t_{50}$. The energy-dependent time of arrival of each photon is taken into account, see section 3.2.

6. Each count in the simulated TGF is randomly extracted from the data set of simulated events obtained in step 2 as seen by the detector before the effects of the electronics are taken into account

7. The hit stream is then processed by the FEE simulator, resulting in the list of counts (time, energy, and detectors hit) as it would be measured by the real detector.

Figure 5 shows a schematic representation of the simulation flow described above, evidencing the main processes affecting the observations. This simulation flow maps the TGF morphology space ($F \times t_{50}$) into the TGF observed space ($N^{\text{obs}} \times t_{50}^{\text{obs}}$), where $N^{\text{obs}}$ is the observed number of counts and $t_{50}^{\text{obs}}$ is the observed duration.

### 3.2. Energy Versus Time Distributions at Satellite Altitude

Simulations presented in previous works (Marisaldi et al., 2014, 2015) assumed photon energy and arrival time at satellite as two independent variables. In addition, photon energy was sampled from an empirical model of the cumulative TGF spectrum. The first assumption is not correct because high-energy photons have a lower probability of scattering in the atmosphere than lower energy photons. Therefore, a very short photon pulse at the source would result in an asymmetric time profile at satellite altitude, with photons in the tens of megaelectronvolts range basically mapping the source time profile, and a delayed tail of Compton-scattered photons in the energy range below a few megaelectronvolts. When pileup is an issue, as in this case, taking into account the correct energy versus time profile of photons is mandatory.

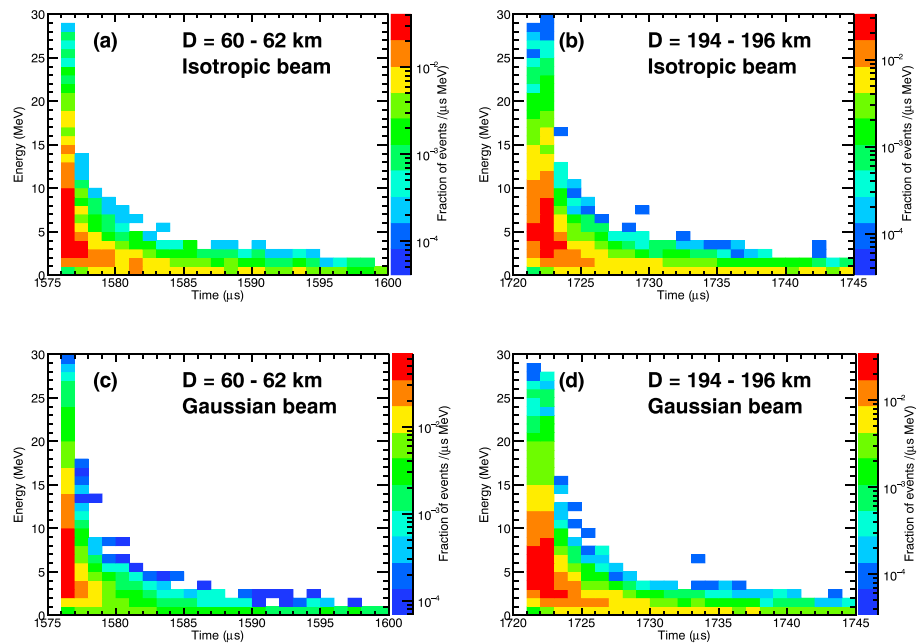

**Figure 6.** Energy versus time of arrival distribution at satellite altitude for a RREA photon spectrum. The RREA is produced at 15-km altitude and propagated through the atmosphere for different source emission geometries and radial distance $D$ from the subsatellite point. RREA = Relativistic Runaway Electron Avalanche.

We addressed this issue by dedicated simulations based on the Geant4 toolkit. Geant4 is developed by the European Organization for Nuclear Research (CERN) in association with a worldwide collaboration. The ability of Geant4 to accurately simulate particle propagation for high-energy radiation in the atmosphere was extensively tested in Rutjes et al. (2016). For these simulations we used the Option4 electromagnetic model. The atmosphere is simulated between 0- and 150-km altitude, and neglected above. The air density/altitude profile follows the NRLMSISE-00 model (Picone et al., 2002). The simulation starts from a photon point source with an energy spectrum proportional to $\exp(-\epsilon/7.3\,\mathrm{MeV})/\epsilon$ and maximum photon energy 30 MeV. The photons are emitted from 15-km altitude, and recorded at 450 km, at several radial distances $D$ (the distance between the subsatellite point and the TGF source footprint). Since it is impossible to record particles at the very exact value of $D$, a small integration interval around it is used, which must be less than 2 km to avoid an artificial broadening of the time distributions of the recorded particles. Two types of beaming were tested: an isotropic cone with an opening half angle of $\theta = 40°$ and a Gaussian beam with a $\sigma_\theta = 15°$. Figure 6 shows the resulting energy versus time of arrival distribution for the two photon production angular distributions at source and two distance $D$ from satellite footprint, the latter corresponding to those for events #10 and #14, respectively (see Table 1). These results were also confirmed by custom-built software used in previous studies (Østgaard et al., 2008). Since there are no significant differences in the energy versus time distributions at satellite altitude between isotropic and Gaussian angular distributions at source, for the simulations shown in this paper we have always used the isotropic angular distribution.

For these simulations we assumed an instantaneous photon production at source. For practical purposes this means that the production time is shorter than the integration time constant of the instrument, that is, about 4 μs in the case of MCAL. When a finite $t_{50}$ is assumed in simulation step 3 we start with the time profile for an instantaneous source and then smooth the arrival time at satellite by a time interval extracted at random by a Gaussian time profile distribution having the corresponding $\sigma$. For $\sigma$ values above few tens of microseconds the original energy dependence of the time profile is smoothed out. However, since many observed high-energy counts belong to short-duration TGFs, this effect must be taken into account. At this stage we are still assuming that the time profile of a TGF consists of a single pulse. We discuss the possibility of a more complex structure of the TGF time profile in section 4.4.

We also point out that, since we know the incoming direction for each event in the high-energy sample, we perform the Monte Carlo simulations assuming an input parallel plane wave of photons from the incoming direction corresponding to the specific TGF under test. This makes the effects of the counts topology (number of detectors hit, energy deposition) more realistic.

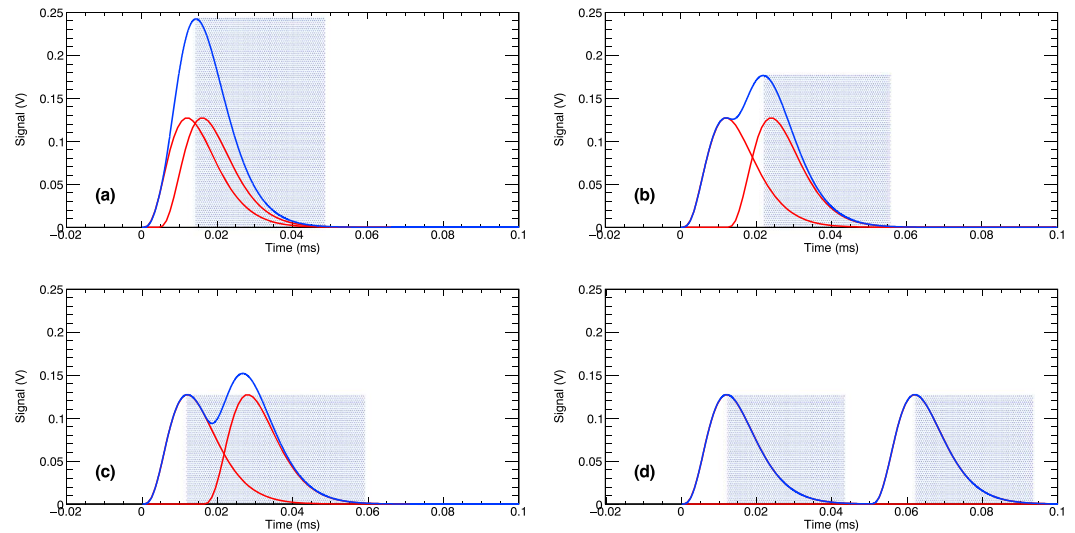

**Figure 7.** Simulations of the signals from one of the front-end electronic chains following two consecutive hits, for different time separation between the two hits (from a to d: 4 μs, 12 μs, 16 μs, and 50 μs, respectively). Red curve: single hit signal output from the shaper amplifier. Blue curve: total signal. Shaded region: trigger pulse.

### 3.3. FEE Model

MCAL includes 30 independent detectors (four of them are permanently disabled due to high electronic noise). Each detector consists of a CsI(Tl) scintillation bar read out by two large-area silicon photodiodes, one at each edge of the bar, each connected to an analog FEE readout chain including charge-sensitive amplifier, a shaper amplifier, a zero-crossing discriminator, and a sample-and-hold stage. Excluding the eight chains serving the disabled detectors, the system consists of 52 identical and independent analog readout chains active at any time. A detailed description of the MCAL instrument is reported in Labanti et al. (2009). A trigger is issued independently for each bar based on the sum of the signals at both bars' ends. We developed a model of the analog FEE based on PSpice (Personal Simulation Program with Integrated Circuit Emphasis) simulations of the actual circuits. Figure 7 shows different working regimes for a single detector readout chain, depending on the time separation $\Delta t$ between the input signals:

1. $\Delta t \leq 16$μs: single trigger on the trailing pulse, the measured amplitude depends on both signals amplitude and time separation (Figures 7a and 7b).
2. $16 \leq \Delta t \leq 40$μs: single trigger on the leading pulse, the measured amplitude is the correct amplitude of the leading signal (Figure 7c).
3. $\Delta t \geq 40$μs: two triggers are issued, both correct amplitudes are collected (Figure 7d).

The boundaries at 16 and 40 μs between the three operational regimes are also dependent on signal amplitude. It is clear that dead-time and pileup effects combine in a complex fashion based on photons time of arrival and energy. In particular, regime 1 can lead to the collection of high-amplitude signals starting from the combination of many relatively low amplitude signals. We target this regime as a viable way to artificially boost to high energies the measured spectrum under conditions of high count rates, as in the case of a bright short TGF. Since the electronics chain can trigger again, that is, the hold signal is released, only when the signal goes below a threshold, an arbitrarily long train of closely spaced pulses can keep the hold signal always active, that is, each individual detector is paralizable.

Once a trigger is generated by one detector, the trigger signal is sent to the digital FEE that handles time stamping and data acquisition. When a trigger is received by the digital FEE, a 2-μs-long coincidence window is opened: all other triggers collected during this window will be regarded as belonging to the initial trigger and will be formatted in a single count with a unique time stamp. The rationale for this is to account for the time jitter and slightly different time constants of different electronic chains. However, in case of high count rate this can result in the incorrect grouping into a single count of hits physically belonging to different events, further boosting to high energies the measured spectrum. After the 2 μs coincidence window, there is a 1-μs-long blanking time window during which no triggers are accepted and the configuration of the triggered bars is stored. Given the 1-MHz clock of the system, the combination of coincidence and blanking time windows makes it impossible to observe counts with time separation shorter than typically 4 μs. This artificially limits the maximum detectable count rate to 250 kHz, which is more than adequate for detecting cosmic gamma ray bursts (GRBs), that was the primary scientific target for MCAL in the design phase but results to be marginal in case of TGFs. We stress that, while the three working regimes described above act asynchronously and independently on each of the 26 active detectors, the counts grouping applies to the instrument as a whole, as well as the consequent minimum time separation of 4 μs between consecutive counts. The FEE model used in this work accounts for all the features presented above, starting from the analog signals in each electronics chain.

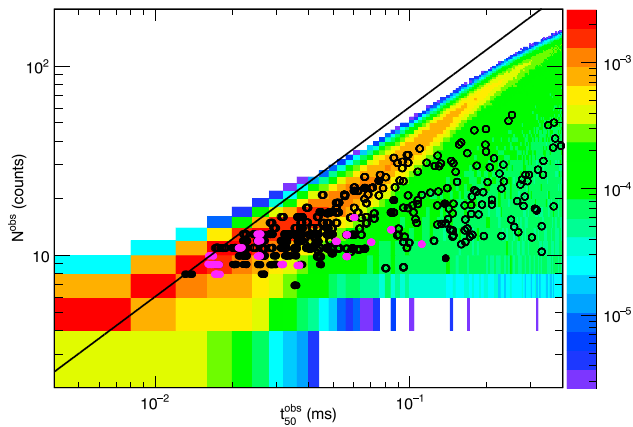

**Figure 8.** HE sample (magenta filled circles) and the rest of the WI sample (black filled circles) in the observed ($N^{obs} \times t^{obs}_{50}$) parameter space. The SC data set (black hollow circles) is shown for reference. The color map shows how a uniform distribution of TGFs in the ($F \times t_{50}$) parameter space is mapped onto the observed ($N^{obs} \times t^{obs}_{50}$) space. Unit is the fraction of simulated TGFs per (4 μs × 2 counts) bin. The simulations correspond to the interaction geometry of event #10, and the total number of simulated events is $10^6$. The black line shows the maximum number of counts allowed by the 250-kHz rate limit set by the digital FEE. HE = WWLLN-identified events with high-energy (>30 MeV) counts; WWLLN = World-Wide Lightning Location Network; WI = WWLLN-identified; SC = selection criteria; TGFs = terrestrial gamma ray flashes; FEE = front-end electronics.

## 4. Discussion

### 4.1. Implications on AGILE TGF Detection Rate

Figure 3 clearly shows that the selection criteria implemented up to the SC sample (Marisaldi et al., 2015) fail to identify about 50% of the TGF associated with WWLLN sferics. This loss of sensitivity is mainly due to the cut on maximum count energy, which must be definitely modified in the forthcoming TGF searches. This cut effectively resulted also in an efficient cosmic ray rejection, which otherwise would appear as a constant background contamination in the longitude and local time distribution of the TGF candidate sample. Therefore, relaxing the cut on maximum count energy must be accompanied by an additional rejection criteria for cosmic ray showers. In Marisaldi et al. (2015) we estimated the AGILE yearly TGF detection rate to be ≈1,000 TGFs/year. If the above mentioned undetected fraction holds for all TGFs and not only for WWLLN-associated events, we may expect the detection of up to ≈2,000 TGFs/year.

### 4.2. TGFs in the (Fluence × Duration) Parameter Space

Figure 8 shows the WI (WWLLN-identified) TGF sample in the observed number of counts—duration ($N^{obs} \times t^{obs}_{50}$) parameter space (filled circles). The SC (selection criteria) sample is shown as a reference (hollow circles). The HE (high-energy) sample (magenta filled circles) apparently clusters at lower number of counts, lower duration, than the low-energy part of the WI sample (black filled circles). The color map refers to a simulated data set for interaction geometry corresponding to event #10 and shows how a uniform distribution of TGFs in the ($0.05$ cm$^{-2}$ < $F$ < $1.95$ cm$^{-2}$) ×($0.001$ ms < $t_{50}$ < $0.5$ ms) parameter space is mapped onto the observed ($N^{obs} \times t^{obs}_{50}$) space. Although this is a special case, and the assumed source distribution is not realistic, this allows us to identify several relevant features in the plot. There are two regions in the observed space that are not accessible, that is, TGFs cannot be observed with certain combinations of duration and number of counts. The bottom right portion of the plot corresponds to long-duration low-fluence TGFs. The boundary at 8–10 counts is due to the minimum fluence (0.05 cm$^{-2}$) chosen for simulations. We set this value simply because of the sensitivity of MCAL and the selection criteria settings will prevent the detection of TGFs with lower fluence. More insightful is the forbidden zone for short-duration high-fluence, that is, high flux, TGFs shown in the top left part of the plot. Because of the combination of dead time and pileup, these events are observed as if they had a much lower fluence. The black line shows the maximum number of counts allowed by the 250-kHz detection rate limit set by the digital FEE. The line slope is multiplied by a factor 2.44 to account for the ratio between $t_{90}$ and $t_{50}$ for a Gaussian pulse. This detected flux limit comes from the hardware limit set by the digital FEE implementation, that joins together in a single count all counts from all detectors collected within a 4-μs time window. The population of short TGFs with $t_{50} \leq 20$μs is very well bound by this line. For longer durations other effects, presumably dead time and pileup, dominate the distribution, and the maximum observed number of counts are lower than the prediction from this maximum rate limit. This is also due to the maximum fluence (1.95 cm$^{-2}$) simulated in this work.

In Marisaldi et al. (2015) we suggested a physical origin for the $N^{obs}$ versus $t_{50}^{obs}$ behavior of the TGF sample, based on our previous understanding of the MCAL dead time. We revise this statement in view of our current understanding of the instrument behavior. The observed sample fills almost perfectly into the observational parameter space allowed by the instrument electronics. This means that the sample is basically shaped by the instrument characteristics and we cannot state any intrinsic property of the TGF sample, at least for durations lower than ~100 µs.

We point out that this bias due to instrumental effect would affect also the observed intensity distribution. In Marisaldi et al. (2014) we obtained an intensity distribution well described by a power law with exponent $\lambda = -2.4$. This is remarkably compliant with the results obtained independently for RHESSI (Østgaard et al., 2012) and Fermi (Tierney et al., 2013) TGFs. However, we point out that that result was obtained for a TGF sample biased toward long durations (median duration 290 µs) because it was observed before the inhibition of the anticoincidence shield described in Marisaldi et al. (2015). Looking at the long duration part of the TGF sample shown in Figure 8, we note that $N^{obs}$ is not clustering at the edge of the permitted zone; therefore, we might expect that the intensity distribution derived from this part of the sample is less affected by instrumental effects. For this reason we can still consider valid the results on the intensity distribution presented in Marisaldi et al. (2014). Conversely, assessing the true intensity distribution for the TGF population with duration shorter than 100 µs is not straightforward and will require additional work.

### 4.3. TGF High-Energy Spectrum

The primary goal of the simulation framework described in section 3 was to understand whether a classical RREA spectrum could be responsible for the observed high-energy counts. We target case study events #10, #14, and #24 introduced in section 2, with maximum observed energy $E_{MAX}$ of 113, 67, and 39 MeV, respectively. Figure 9 shows color maps of the regions in the $(F \times t_{50})$ parameter space resulting in at least one observed count with energy in the interval $E_{MAX} \pm 20\%$. This margin accounts for the expected MCAL energy resolution and systematic error on energy reconstruction in the tens of megaelectronvolt range. The shape of these regions shows that the critical parameter is the total flux, for which a proxy is the ratio between fluence and duration. In other words it is possible to obtain high-energy counts either with a relatively low fluence for a short duration or with higher fluence for a longer duration. The permitted region extends toward longer duration if the requirement on $E_{MAX}$ is lower, see the trend from Figure 9a to Figure 9c.

When, in addition to the requirement on $E_{MAX}$, we also require duration and observed number of counts to be compliant to observations, we obtain the contour plots overplotted on the color maps in Figure 9. Here we assumed a 20% margin on duration and 30% margin on the number of counts, corresponding to the average relative error ($1\sigma$) resulting by the maximum likelihood technique described previously. Contour levels (innermost to outermost) indicates the 25%, 50%, 75%, and 95% percentiles of the total number of counts. Figure 9a shows that, in order to observe an event like #10, with 90 MeV $< E_{MAX} <$ 136 MeV the original fluence must be larger than 0.42 cm$^{-2}$ (95% confidence). In order to observe an event like #14, with 54 MeV $< E_{MAX} <$ 80 MeV the original fluence must be larger than 0.24 cm$^{-2}$ (95% confidence). Tierney et al. (2013) reports 10 TGFs with fluence larger than 0.20 cm$^{-2}$ and one with fluence larger than 0.35 cm$^{-2}$, out of a representative sample of 106 TGFs detected by Fermi. We observe nine events out of 79 with maximum energy larger then 60 MeV (11.4% of the sample), roughly in agreement with the 9.4% of the Fermi sample with fluence larger than 0.2 cm$^{-2}$. However, we cannot push this comparison further because of the differences in the selection criteria applied for the two samples, our criterion in particular being based on association with WWLLN sferics, which bias the sample toward shorter, and possibly more energetic, TGFs.

Figure 9c shows the interesting case of a TGF (event #24) with $E_{MAX} = 39$ MeV, a value close to the maximum expected from a RREA spectrum. Here we can see how the 95% contour plot bends toward a region of relatively low fluences and longer durations, which is less affected by instrumental effects. This means that the observed 39-MeV count can be the true energy of a single photon. However, the core region of the $(F \times t_{50})$ parameter space compatible with observations still corresponds to short durations and very high fluences, whose related observations are likely dominated by pileup and dead time. These conclusions are based on the assumption of a source emitting on a uniform wide beam of 40° opening half angle. However, for the considered case studies, the photon energy versus arrival time distribution is not significantly different for the two source emission geometry models considered (see section 3.2); therefore, we regard these conclusions independent on the assumed source model. Other case studies, with larger source distances to the satellite footprint, would be needed to test the source geometry with this approach.

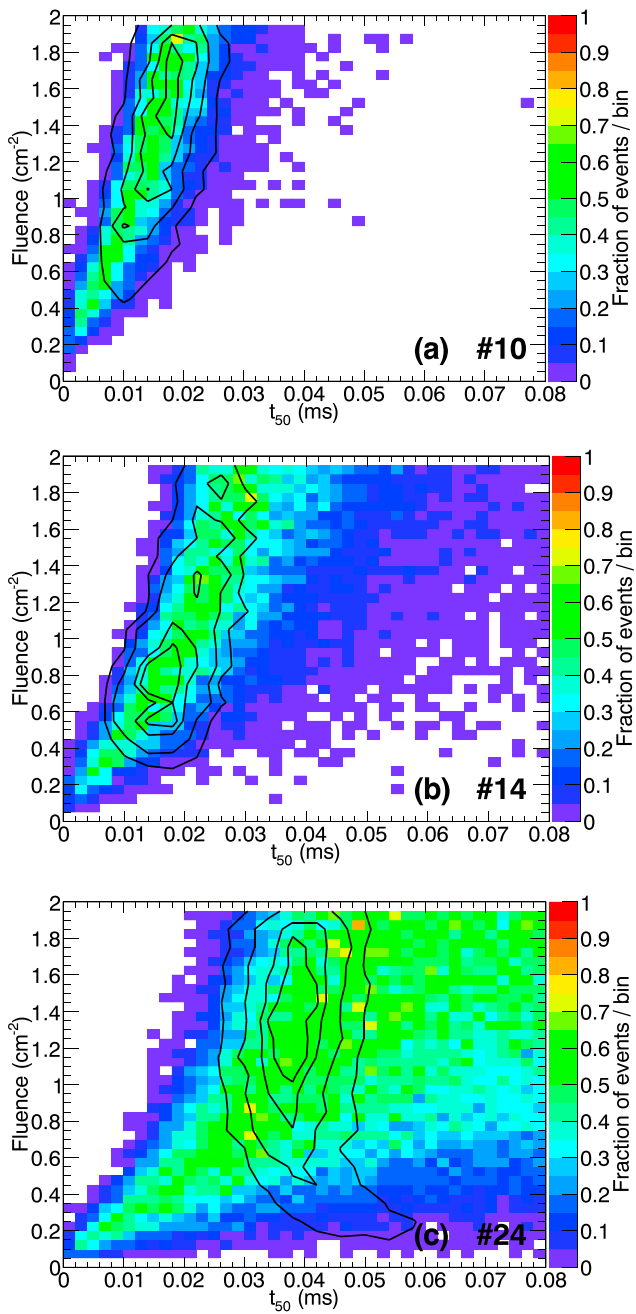

**Figure 9.** Each plot corresponds to a case study event: (a) #10, (b) #14, and (c) #24. Color plot: fraction of simulated TGFs in the ($F \times t_{50}$) parameter space resulting in at least one observed count with energy in the interval $E_{MAX} \pm 20\%$. Contour plot: portion of the simulated data set resulting in $N^{obs}$ and $t^{obs}_{50}$ compliant with observations, allowing for a ±20% uncertainty on duration and a ±30% uncertainty on the number of counts. Contour levels (innermost to outermost) indicates the 25%, 50%, 75%, and 95% percentiles. TGFs = terrestrial gamma ray flashes.

### 4.4. TGF Fine Time Structure at the Source

In the previous section we assumed a RREA spectrum produced in a single Gaussian-shaped time profile and identified the region in the source parameter space that is compatible with observations. In order to justify the observed maximal count energies and event duration, we need fluences at the highest edge of the fluence distribution observed so far. If we consider the lowest fluences consistent with the allowed regions in Figure 9 the rate of events with high-energy counts is marginally compatible with expectations from the intensity distribution from Fermi. However, we can relax the requirement on fluence if we allow the pulse duration to be shorter, see the color maps in Figure 9. If this were the case, the time profile of a TGF at source could be the superposition of several quasi-instantaneous pulses, each of them reflecting an avalanche process, whose envelop is detected at satellite altitude as a single pulse because of lack of counts statistics and time smearing due to Compton scattering. Pulse superposition was presented in Briggs et al. (2010) and Fishman et al. (2011) and extensively discussed in Celestin and Pasko (2012) as a possible explanation for long-duration TGFs. In particular, Celestin and Pasko (2012) suggest that a single instantaneous pulse at source is compatible with the shortest TGF durations observed at satellite altitude so far (few tens of microseconds). This result is supported also by our simulations shown in Figure 6. However, another study (Fitzpatrick et al., 2014) reports that a single instantaneous pulse is not capable to explain the spectral behavior of TGFs detected by Fermi and suggests that an intrinsic source pulse duration of at least few tens of microseconds is necessary. Magnetic field measurements reported in Cummer et al. (2011) are indicative of a current component mimicking the gamma ray light curve of a TGF detected by Fermi supporting the scenario of a long-duration (tens of microseconds) production process at the source.

In case of a pulse at the satellite with duration shorter than the integration time constant of the detector (≈4 µs) we expect MCAL to detect a single count with total reconstructed energy dependent on the pulse fluence. Given the few microseconds spread due to photon transport through the atmosphere, this pulse could correspond to a much shorter (submicrosecond) photon burst at the source. We investigated the possibility for such pulses by searching the high-energy data set for counts with $E > 30$ MeV and with a time difference with respect to the previous count larger than 10 µs and lower than 200 µs. The minimum time difference is set in order to be sure that there is a real time separation between consecutive counts and the effect cannot be ascribed to the counts grouping performed at the digital FEE level (see section 3.3). The maximum time difference is set in order to be reasonably sure that the count is associated to the TGF and not just a background count. We found three events out of 22 (the total number of events in the high-energy sample, excluding the spurious events) with counts satisfying those conditions. Given the limited statistics, we cannot exclude that these events are due to chance association to background high-energy counts, as discussed in section 2. Given the background rate above 30 MeV we can expect to have on average two high-energy background counts in the high-energy sample, excluding the spurious events, the probability of having three being a nonnegligible value of 0.19. Therefore, we conclude that the presented data do not show any clear evidence of a fine time structure of TGFs.

Recently, a method has been proposed (Dwyer & Cummer, 2013; Mezentsev et al., 2017) to investigate the TGF fine time structure at the source based on the spectral characteristics of the associated very low frequency (VLF) sferics. We searched for VLF measurements associated to the high-energy sample from the sensors managed by the Duke University (North Carolina, USA), but no good associations were found. This search should be repeated with a larger sample.

## 5. Summary and Conclusions

This study shows that it is possible to explain the observed high-energy counts by a classical RREA production spectrum, after the correct energy-dependent photon transport through the atmosphere and the detailed model of the MCAL FEE are properly taken into account. Data for the WWLLN-identified data set can be accessed at this URL (http://www.ssdc.asi.it/mcalwtgfcat/). The fraction of events with high-energy counts is roughly in agreement with the fraction of high-fluence ($F > 0.2$ cm$^{-2}$) events detected by Fermi although the limited size of the samples and the different selection criteria may affect this result. Although RREA provides an acceptable explanation to the AGILE observations, solving a 7-year-old controversy, it cannot be ruled out that a deviation from the RREA spectrum at high energy may still exist. In other words, a harder spectrum than RREA as input to the simulations described in section 3 could still provide an acceptable solution region in Figure 9. This possibility can be explored further by simulations, but high-quality measurements in the tens of megaelectronvolt regime are needed. AGILE Gamma Ray Imaging Detector data can be exploited further for this purpose. Concerning the fine time structure of TGFs, our measurements do not show any clear evidence of a fine time structure of TGFs on microsecond time scale. Finally, valuable insight both on the TGF high-energy spectrum and time structure will be provided by the Modular X- and Gamma ray Sensor (MXGS) of the recently launched ASIM mission (Østgaard et al., 2019), sensitive up to 40 MeV and with a readout electronics specifically designed and tailored for the first time to an accurate control of dead time and pileup effects in the TGF high-photon flux regime.

**Acknowledgments**

AGILE is a mission of the Italian Space Agency (ASI), with coparticipation of INAF (Istituto Nazionale di Astrofisica) and INFN (Istituto Nazionale di Fisica Nucleare). We acknowledge partial support through ASI Grant I/028/12/2. This study was supported by the European Research Council under the European Unions Seventh Framework Programme (FP7/2007-2013)/ERC Grant Agreement 320839 and the Research Council of Norway under contracts 208028/F50 and 223252/F50 (CoE). Some part of the simulations were performed on resources provided by UNINETT Sigma2—the National Infrastructure for High Performance Computing and Data Storage in Norway, under Project NN9526K. The authors wish to thank the World Wide Lightning Location Network (http://wwlln.net), a collaboration among over 50 universities and institutions, for providing the lightning location data used in this paper. The authors thank Paolo Bastia at Thales Alenia Space Italia for his fundamental contribution to the development of the MCAL front-end electronics numerical model. M. Marisaldi wishes to thank the Fulbright Research Scholar Program. The TGF data presented in this work are publicly available at the ASI Space Science Data Center (SSDC) website (http://www.ssdc.asi.it/mcalwtgfcat/). The MCAL FEE simulation software used in this work is publicly available at this website (https://github.com/martinomaris/MCAL-front-end-simulator).

## References

Agostinelli, S., Allison, J., Amako, K., Apostolakis, J., Araujo, H., Arce, P., et al. (2003). Geant4—a simulation toolkit. *Nuclear Instruments and Methods in Physics Research Section A*, *506*(3), 250–303. https://doi.org/10.1016/S0168-9002(03)01368-8

Albrechtsen, K. H., Ostgaard, N., Berge, N., & Gjesteland, T. (2019). Observationally weak TGFs in the RHESSI data. *Journal of Geophysical Research: Atmospheres*, *124*, 287–298. https://doi.org/10.1029/2018JD029272

Allison, J., Amako, K., Apostolakis, J., Araujo, H., Dubois, P. A., Asai, M., et al. (2006). Geant4 developments and applications. *IEEE Transactions on Nuclear Science*, *53*, 270–278. https://doi.org/10.1109/TNS.2006.869826

Babich, L. P., Bochkov, E. I., Kutsyk, I. M., & Rassoul, H. K. (2014). Analysis of fundamental interactions capable of producing neutrons in thunderstorms. *Physical Review D*, *89*(9), 93010. https://doi.org/10.1103/PhysRevD.89.093010

Babich, L. P., & Roussel-Dupre, R. A. (2007). Origin of neutron flux increases observed in correlation with lightning. *Journal of Geophysical Research*, *112*, D13303. https://doi.org/10.1029/2006JD008340

Bowers, G. S., Smith, D. M., Martinez-McKinney, G. F., Kamogawa, M., Cummer, S. A., Dwyer, J. R., & Kawasaki, Z. (2017). Gamma ray signatures of neutrons from a terrestrial gamma ray flash. *Geophysical Research Letters*, *44*, 10,063–10,070. https://doi.org/10.1002/2017GL075071

Briggs, M. S., Fishman, G. J., Connaughton, V., Bhat, P. N., Paciesas, W. S., Preece, R. D., & Chekhtman, A. (2010). First results on terrestrial gamma ray flashes from the Fermi Gamma-ray Burst Monitor. *Journal of Geophysical Research*, *115*, 7323. https://doi.org/10.1029/2009JA015242

Carlson, B. E., Lehtinen, N. G., & Inan, U. S. (2010). Neutron production in terrestrial gamma ray flashes. *Journal of Geophysical Research*, *115*, A00E19. https://doi.org/10.1029/2009JA014696

Celestin, S., & Pasko, V. P. (2012). Compton scattering effects on the duration of terrestrial gamma-ray flashes. *Geophysical Research Letters*, *39*, 2802. https://doi.org/10.1029/2011GL050342

Celestin, S., Xu, W., & Pasko, V. P. (2012). Terrestrial gamma ray flashes with energies up to 100 MeV produced by nonequilibrium acceleration of electrons in lightning. *Journal of Geophysical Research*, *117*, 5315. https://doi.org/10.1029/2012JA017535

Celestin, S., Xu, W., & Pasko, V. P. (2015). Variability in fluence and spectrum of high-energy photon bursts produced by lightning leaders. *Journal of Geophysical Research: Space Physics*, *120*, 10,712–10,723. https://doi.org/10.1002/2015JA021410

Chilingarian, A., Hovsepyan, G., & Kozliner, L. (2013). Thunderstorm ground enhancements: Gamma ray differential energy spectra. *Physics Review D*, *88*, 73001. https://doi.org/10.1103/PhysRevD.88.073001

Connaughton, V., Briggs, M. S., Holzworth, R. H., Hutchins, M. L., Fishman, G. J., Wilson-Hodge, C. A., & Smith, D. M. (2010). Associations between Fermi Gamma-ray Burst Monitor terrestrial gamma ray flashes and sferics from the World Wide Lightning Location Network. *Journal of Geophysical Research*, *115*, 12307. https://doi.org/10.1029/2010JA015681

Connaughton, V., Briggs, M. S., Xiong, S., Dwyer, J. R., Hutchins, M. L., Grove, J. E., & Wilson-Hodge, C. (2013). Radio signals from electron beams in terrestrial gamma ray flashes. *Journal of Geophysical Research: Space Physics*, *118*, 2313–2320. https://doi.org/10.1029/2012JA018288

Cummer, S. A., Lu, G., Briggs, M. S., Connaughton, V., Xiong, S., Fishman, G. J., & Dwyer, J. R. (2011). The lightning-TGF relationship on microsecond timescales. *Geophysical Research Letter*, *38*, 14810. https://doi.org/10.1029/2011GL048099

Dwyer, J. R. (2003). A fundamental limit on electric fields in air. *Geophysical Research Letters*, *30*(20), 2055. https://doi.org/10.1029/2003GL017781

Dwyer, J. R. (2012). The relativistic feedback discharge model of terrestrial gamma ray flashes. *Journal of Geophysical Research, 117*, 2308. https://doi.org/10.1029/2011JA017160

Dwyer, J. R., & Cummer, S. A. (2013). Radio emissions from terrestrial gamma-ray flashes. *Journal of Geophysical Research: Space Physics, 118*, 3769–3790. https://doi.org/10.1002/jgra.50188

Dwyer, J. R., & Smith, D. M. (2005). A comparison between Monte Carlo simulations of runaway breakdown and terrestrial gamma-ray flash observations. *Geophysical Research Letter, 32*, 22804. https://doi.org/10.1029/2005GL023848

Dwyer, J. R., Smith, D. M., & Cummer, S. A. (2012). High-energy atmospheric physics: terrestrial gamma-ray flashes and related phenomena. *Space Science Reviews, 173*, 133–196. https://doi.org/10.1007/s11214-012-9894-0

Enoto, T., Wada, Y., Furuta, Y., Nakazawa, K., Yuasa, T., Okuda, K., & Tsuchiya, H. (2017). Photonuclear reactions triggered by lightning discharge. *Nature, 551*(7681), 481. https://doi.org/10.1038/nature24630

Fishman, G. J. (1994). Discovery of intense gamma-ray flashes of atmospheric origin. *Science, 264*, 1313–1316.

Fishman, G. J., Briggs, M. S., Connaughton, V., Bhat, P. N., Paciesas, W. S., von Kienlin, A., & Greiner, J. (2011). Temporal properties of the terrestrial gamma-ray flashes from the Gamma-Ray Burst Monitor on the Fermi Observatory. *Journal of Geophysical Research, 116*, 7304. https://doi.org/10.1029/2010JA016084

Fitzpatrick, G., Cramer, E., McBreen, S., Briggs, M. S., Foley, S., Tierney, D., & von Kienlin, A. (2014). Compton scattering in terrestrial gamma-ray flashes detected with the Fermi gamma-ray burst monitor. *Physical Review D, 90*(4), 43008. https://doi.org/10.1103/PhysRevD.90.043008

Gjesteland, T., Østgaard, N., Connell, P. H., Stadsnes, J., & Fishman, G. J. (2010). Effects of dead time losses on terrestrial gamma ray flash measurements with the Burst and Transient Source Experiment. *Journal of Geophysical Research, 115*, A00E21. https://doi.org/10.1029/2009JA014578

Grefenstette, B. W., Smith, D. M., Hazelton, B. J., & Lopez, L. I. (2009). First RHESSI terrestrial gamma ray flash catalog. *Journal Geophysical Research, 114*, 2314. https://doi.org/10.1029/2008JA013721

Gurevich, A. V., Milikh, G. M., & Roussel-Dupre, R. (1992). Runaway electron mechanism of air breakdown and preconditioning during a thunderstorm. *Physics Letters A, 165*, 463–468. https://doi.org/10.1016/0375-9601(92)90348-P

Hazelton, B. J., Grefenstette, B. W., Smith, D. M., Dwyer, J. R., Shao, X. M., Cummer, S. A., & Holzworth, R. H. (2009). Spectral dependence of terrestrial gamma-ray flashes on source distance. *Geophysical Research Letter, 36*, L01108. https://doi.org/10.1029/2008GL035906

Koshut, T. M., Paciesas, W. S., Kouveliotou, C., van Paradijs, J., Pendleton, G. N., Fishman, G. J., & Meegan, C. A. (1996). Systematic effects on duration measurements of gamma-ray bursts. *The Astrophysical Journal, 463*, 570. https://doi.org/10.1086/177272

Labanti, C., Marisaldi, M., Fuschino, F., Galli, M., Argan, A., Bulgarelli, A., & Trifoglio, M. (2009). Design and construction of the mini-calorimeter of the AGILE satellite. *Nuclear Instruments and Methods in Physics Research A, 598*, 470–479. https://doi.org/10.1016/j.nima.2008.09.021

Luque, A. (2014). Relativistic runaway ionization fronts. *Physical Review Letters, 112*(4), 045003. https://doi.org/10.1103/PhysRevLett.112.045003

Mailyan, B. G., Briggs, M. S., Cramer, E. S., Fitzpatrick, G., Roberts, O. J., Stanbro, M., & Dwyer, J. R. (2016). The spectroscopy of individual terrestrial gamma-ray flashes: Constraining the source properties. *Journal of Geophysical Research: Space Physics, 121*, 11,346–11,363. https://doi.org/10.1002/2016JA022702

Marisaldi, M., Argan, A., Trois, A., Giuliani, A., Tavani, M., Labanti, C., & Salotti, L. (2010). Gamma-ray localization of terrestrial gamma-ray flashes. *Physical Review Letters, 105*(12), 128501. https://doi.org/10.1103/PhysRevLett.105.128501

Marisaldi, M., Argan, A., Ursi, A., Gjesteland, T., Fuschino, F., Labanti, C., & Trois, A. (2015). Enhanced detection of terrestrial gamma-ray flashes by AGILE. *Geophysical Research Letters, 42*, 9481–9487. https://doi.org/10.1002/2015GL066100

Marisaldi, M., Fuschino, F., Labanti, C., Galli, M., Longo, F., Del Monte, E., et al. (2010). Detection of terrestrial gamma ray flashes up to 40 MeV by the AGILE satellite. *Journal of Geophysical Research, 115*, A00E13. https://doi.org/10.1029/2009JA014502

Marisaldi, M., Fuschino, F., Tavani, M., Dietrich, S., Price, C., Galli, M., et al. (2014). Properties of terrestrial gamma ray flashes detected by AGILE MCAL below 30 MeV. *Journal of Geophysical Research: Space Physics, 119*, 1337–1355. https://doi.org/10.1002/2013JA019301

Mezentsev, A., Lehtinen, N., Østgaard, N., Pérez-Invernón, F. J., & Cummer, S. A. (2017). Spectral characteristics of VLF Sferics associated with RHESSI TGFs. *Journal of Geophysical Research: Atmospheres, 123*, 139–159. https://doi.org/10.1002/2017JD027624

Østgaard, N., Albrechtsen, K. H., Gjesteland, T., & Collier, A. (2015). A new population of terrestrial gamma-ray flashes in the RHESSI data. *Geophysical Research Letters, 42*, 10,937–10,942. https://doi.org/10.1002/2015GL067064

Østgaard, N., Balling, J. E., Bjørnsen, T., Brauer, P., Budtz-Jørgensen, C., Bujwan, W., & Yang, S. (2019). The Modular X- and Gamma-Ray Sensor (MXGS) of the ASIM payload on the International Space Station. *Space Science Reviews, 215*(2), 23. https://doi.org/10.1007/s11214-018-0573-7

Østgaard, N., Gjesteland, T., Hansen, R. S., Collier, A. B., & Carlson, B. (2012). The true fluence distribution of terrestrial gamma flashes at satellite altitude. *Journal of Geophysical Research, 117*, 3327. https://doi.org/10.1029/2011JA017365

Østgaard, N., Gjesteland, T., Stadsnes, J., Connell, P. H., & Carlson, B. (2008). Production altitude and time delays of the terrestrial gamma flashes: Revisiting the Burst and Transient Source Experiment spectra. *Journal Geophysical Research, 113*, 2307. https://doi.org/10.1029/2007JA012618

Picone, J. M., Hedin, A. E., Drob, D. P., & Aikin, A. C. (2002). NRLMSISE-00 empirical model of the atmosphere: Statistical comparisons and scientific issues. *Journal of Geophysical Research, 107*, 1468. https://doi.org/10.1029/2002JA009430

Rutjes, C., Sarria, D., Broberg Skeltved, A., Luque, A., Diniz, G., Østgaard, N., & Ebert, U. (2016). Evaluation of Monte Carlo tools for high energy atmospheric physics. *Geoscientific Model Development, 9*, 3961–3974. https://doi.org/10.5194/gmd-9-3961-2016

Smith, D. M. (2005). Terrestrial gamma-ray flashes observed up to 20 MeV. *Science, 307*, 1085–1088.

Smith, D. M., Buzbee, P., Kelley, N. A., Infanger, A., Holzworth, R. H., & Dwyer, J. R. (2016). The rarity of terrestrial gamma-ray flashes: 2. RHESSI stacking analysis. *Journal of Geophysical Research: Atmospheres, 121*, 11,382–11,404. https://doi.org/10.1002/2016JD025395

Tavani, M., Argan, A., Paccagnella, A., Pesoli, A., Palma, F., Gerardin, S., & Giommi, P. (2013). Possible effects on avionics induced by terrestrial gamma-ray flashes. *Natural Hazards and Earth System Sciences, 13*(4), 1127–1133. https://doi.org/10.5194/nhess-13-1127-2013

Tavani, M., Marisaldi, M., Labanti, C., Fuschino, F., Argan, A., Trois, A., & Zanello, D. (2011). Terrestrial gamma-ray flashes as powerful particle accelerators. *Physical Review Letters, 106*(1), 18501. https://doi.org/10.1103/PhysRevLett.106.018501

Tierney, D., Briggs, M. S., Fitzpatrick, G., Chaplin, V. L., Foley, S., McBreen, S., & Wilson-Hodge, C. (2013). Fluence distribution of terrestrial gamma ray flashes observed by the Fermi Gamma-ray Burst Monitor. *Journal of Geophysical Research: Space Physics, 118*, 6644–6650. https://doi.org/10.1002/jgra.50580